\documentclass[11pt]{revtex4}
\usepackage{bm}  
\usepackage{graphicx}
\usepackage{amssymb}
\usepackage{color}
\usepackage{amscd}
\usepackage{epsfig}

\newcommand{\beq}{\begin{equation}}
\newcommand{\eqref}[1]{Eq.\ (\ref{#1})}

\newcommand{\enq}{\end{equation}}

\newcommand{\la}{\langle}
\newcommand{\ra}{\rangle}

\begin{document}

\title{Models of plastic depinning of driven disordered systems}

\author{M. Cristina Marchetti}
\affiliation{Physics Department, Syracuse University, Syracuse NY
13244, USA.}

\date{\today}

\begin{abstract}
Two classes of models of driven disordered systems that exhibit
history-dependent dynamics are discussed. The first class
incorporates local inertia in the dynamics via nonmonotonic stress
transfer between adjacent degrees of freedom. The second class
allows for proliferation of topological defects due to the
interplay of strong disorder and drive. In mean field theory both
models exhibit a tricritical point as a function of disorder
strength. At weak disorder depinning is continuous and the sliding
state is unique. At strong disorder depinning is discontinuous and
hysteretic.
\end{abstract}
\maketitle

\section{Introduction}
Nonequilibrium depinning transitions from static to moving states
underlie the physics of a wide range of phenomena, from fracture
propagation in heterogeneous solids to flux flow in type-II
superconductors \cite{Fisher98}. One class of models, overdamped
{\it elastic} media pulled by a uniform force, has been studied
extensively.  These exhibit a nonequilibrium phase transition from
a pinned  to a sliding state  at a critical value $F_T$ of the
driving force. This nonequilibrium transition displays universal
critical behavior as in equilibrium {\it continuous} transitions,
with the medium's mean velocity $v$ acting as an order parameter
\cite{DSF85,NF92}. In overdamped elastic media, the sliding state
is unique and no hysteresis can occur \cite{AAMunique}.

The elastic medium model is often inadequate to describe real
physical systems that exhibit plastic response on various scales.
Plasticity, used here in a loose sense, may arise in systems with
an underlying periodic structure, such as vortex lattices or
charge density waves, for strong disorder. This yields large
deformations of the driven medium with the proliferation of
topological defects, which are continuously generated and healed
by the interplay of drive, disorder and
interactions.\cite{Bhattacharya93,Nori96,tonomura99} In other
situations, such as crack propagation in heterogeneous
solids\cite{Perrin94}, and the motion of a helium-4 droplet
contact line on a rough surface \cite{Prevost02}, the dissipative
elastic medium model is made inadequate by the presence of
inertial effects or other nonlocal stress propagation mechanisms.
In both cases the depinning transition may become discontinuous
and sometime hysteretic.

Several coarse-grained models of driven extended systems that can
lead to history-dependent dynamics have been proposed in the
literature \cite{Fisher98,strogatz88&89,levy92,MMP2000,SF01}. Here
we examine two classes of such models. In the first class of
models the displacement of the driven medium from some undeformed
reference configuration remains single-valued, as appropriate for
systems without topological defects, but the elastic interactions
are modified by assuming local underdamped dynamics. This is
modeled via a linear stress-strain relation, where the stress
transfer between displacements of different parts of the manifold
is nonmonotonic in time. Models of this type have been used to
incorporate the effect of inertia or elastic stress overshoot in
crack propagation in solids \cite{SF01,SF02}. A related model was
proposed by the author and collaborators as an \emph{effective}
description of topological defects in the driven system
\cite{MMP2000,MCMKD02,MMSS2003}. Below we show that  the presence
of topological defects in a solid can be described as a viscous
force that allows a moving portion of the medium to overshoot a
static configuration before relaxing back to it, so that these two
models are identical. In the second class of models topological
defects are explicitly allowed by removing the constrain that
displacements are single-valued  \cite{strogatz88&89,SSMM04}. We
consider a simple realization of this type of models  proposed
some time ago in the context of charge density waves, where the
scalar displacement describes the phase of the electronic
condensate and the system only exhibits one-dimensional
periodicity along the direction of motion. It is obtained by
assuming a nonlinear coupling among neighboring displacements that
incorporates the crucial feature that the displacement becomes
undefined at the location where the amplitude of the order
parameter collapses. This is incorporated in the model as an
instantaneous jump of the displacement or phase of an amount equal
to its period, knows as phase slip.

The two classes of models exhibit remarkably similar behavior in
the mean-field limit, where the interaction are assumed to have
infinite range. As a function of disorder strength, they both have
a tricritical point separating continuous from discontinuous
depinning transitions.  Depinning is continuous at weak disorder
with mean-field critical exponents that are identical to those
obtained for a dissipative elastic medium. Above a critical
disordered strength, depinning becomes discontinuous and
hysteretic. The origin of the hysteresis is, however, different in
the two models, as discussed below. Preliminary numerical work
also suggests that differences may arise in finite dimensions.

\section{Two classes of models: viscous and phase slip couplings}
We restrict ourselves to the dynamics of a scalar field describing
deformations along the direction of mean motion and discretize
space, denoting by $u_i(t)$ the displacement of the $i$- degree of
freedom from some undeformed reference configuration, at time $t$.
We stress that all models discussed are \emph{coarse-grained}
ones, with $u_i$ representing the displacement of a region pinned
collectively by disorder. The microscopic dynamics is always
assumed to be overdamped, but velocity-dependent couplings can
arise upon coarse-graining. The equation of motion in the
laboratory frame is written by balancing all the forces acting on
each segment of the driven medium as \cite{convective}
\begin{equation}
\label{main_eq}
\partial_tu_i=\sigma^\alpha(\{u_i\},t)+F+f_p(i,u_i)\;,
\end{equation}
where time has been scaled so that the friction coefficient is
unity, $F$ is the external driving force, $f_p$ is the pinning
force, and $\sigma^\alpha$ represents the  stresses due to
interactions with the neighbors. The label $\alpha$ will be used
below to identify two ways of modeling the coupling.  For driven
periodic manifolds, such as vortex lattices or charge density
waves, the pinning force is periodic in the displacement $u_i$,
with $0\leq u_i\leq 1$, and can be written as
$f_p(i,u_i)=h_iY(u_i-\gamma_i)$, where $Y(u)$ is a periodic
function specified below, $h_i$ are random forces with
distribution $\rho(h_i)$, and $\gamma_i$ are random phases chosen
independently and uniformly in $[0,1)$.

First, we consider a class of models with single-valued
displacements $u_i$ and  a linear stress-strain relation of the
general form,
\begin{equation}
\label{stress_strain} \sigma^{\alpha}(\{u_i\},t)=\int_{-\infty}^t
dt'~\frac{1}{Z}\sum_{\la
j\ra}J_{ij}^\alpha(t-t')\big[u_j(t')-u_i(t')\big]\;,
\end{equation}
where the sum is over the $Z$ sites $j$ that are nearest neighbors
of $i$. A suitable choice of the stress-transfer or memory
function, $J_{ij}^\alpha(t)$, yields various models described in
the literature. The familiar elastic model is obtained by assuming
instantaneous stress transfer, i.e., $J_{ij}^{\rm
E}(t)=K\delta(t)$, with $K$ an elastic constant. More generically,
all monotonic models, defined as those with $J_{ij}(t)\geq 0$, for
all $i,j,t$, exhibit a continuous depinning transition with
universal critical behavior and a unique sliding state
\cite{Fisher98}. The choice $J_{ij}^{\rm
V}=\eta\partial_t\delta(t)$ yields purely viscous stresses, i.e.,
$\sigma^{\rm V}=\eta\sum_{\la
j\ra}\big[\dot{u}_j(t)-\dot{u}_i(t)\big]$, provided we identify
$v_i=\dot{u}_i\equiv\partial_t u_i$ with the local flow velocity
of the medium. In this case depinning always occurs at $F=0$ for
pinning force distributions without a finite lower bound, but
there is a critical point above which the system can switch
discontinuously and hysteretically between a macroscopic slow
moving and a fast moving state \cite{MMP2000,MCMKD02}. The author
and collaborators recently considered models where the stress
transfer function is taken to have the form appropriate for a
viscoelastic fluid, that responds elastically on short time scales
and flows viscously at long times \cite{MMP2000}. This
viscoelastic coupling was proposed as an effective way of
describing the local slip due to dislocations generated at the
boundaries between coarse-grained degrees of freedom. The
connection between the viscoelastic model and the presence of free
dislocations in the medium was made more precise in Ref.
\cite{MCMKS02} where it was shown that the equations describing
the dynamics of \emph{equilibrium} deformations of a
two-dimensional lattice with a finite concentration $n_d$ of free
\emph{annealed} dislocations can be recast in the form of the
phenomenological equations of a viscoelastic fluid introduced many
years ago by Maxwell \cite{boonyip80}. In a scalar version of the
equations of viscoelasticity, local compressional stresses are
written in the form given in Eq.~(\ref{stress_strain}), with
\begin{equation}
\label{JVE} J_{ij}^{\rm
VE}(t)=K_\infty\delta(t)-\frac{K_\infty-K_0}{\tau}e^{-t/\tau}\;,
\end{equation}
where $K_\infty$ and $K_0$ are the high and low frequency
compressional moduli, respectively, and $\tau\sim 1/n_d$ is a
microscopic relaxation time \cite{MCMKS02}. Shear stresses have a
similar form, with shear moduli replacing compressional ones and
$G_0=0$ as a fluid has no zero-frequency elastic restoring forces
in response to shear stresses. The nonzero long-wavelength
compressional elasticity ($K_0\not=0$) is associated with density
conservation and plays a crucial role in controlling the physics
of depinning in driven lattices. For this reason in our scalar
model we assume a coupling of the form (\ref{JVE}) in all
directions. On time scales $t$ short compared to $\tau$, the
contribution to the stress coming from the second term in
Eq.~(\ref{JVE}) is negligible compared to the first and
$\sigma^{\rm VE}$ reduces to the stress of an elastic solid,
\begin{equation}
\sigma^{\rm VE}(\{u_i\},t<<\tau)\approx
\frac{K_\infty}{Z}\sum_{\la j\ra}[u_j(t)-u_i(t)]\;.
\end{equation}
Conversely, for $t>>\tau$, one can expand the relative
displacements for $t'\sim t$, and
\begin{eqnarray}
\label{sigmaVE} \sigma^{\rm VE}(\{u_i\},t>>\tau)\approx
\frac{K}{Z}\sum_{\la j\ra}[u_j(t)-u_i(t)]+\frac{\eta}{Z} \sum_{\la
j\ra}[\dot{u}_j(t)-\dot{u}_i(t)]\;,
\end{eqnarray}
with $K=K_0$ and $\eta=(K_\infty-K_0)\tau$. The model of driven
depinning studied below is obtained from Eq.~(\ref{main_eq}) with
the simplified form given in Eq.~(\ref{sigmaVE}) for the stress
strain relation,
\begin{equation}
\label{VE_model} \partial_tu_i=\frac{K}{Z}\sum_{\la
j\ra}[u_j(t)-u_i(t)]+\frac{\eta}{Z} \sum_{\la
j\ra}[\dot{u}_j(t)-\dot{u}_i(t)]+F+f_p(i,u_i)\;,
\end{equation}
and will be referred to as viscous/elastic model (VE).  The
presence  dislocations generated by disorder is incorporated in a
mean-field-type approximation as a local inertial response of the
driven system embodied by the viscous coupling of strength
$\eta\sim 1/n_d$. This model assumes a fixed number of topological
defects and does not describe the creation and annihilation of
dislocations due to the interplay of drive, disorder and
interactions. Furthermore, in a driven disordered solid unbound
dislocations can be pinned by disorder and do not equilibrate with
the lattice. The resulting dynamics cannot be described as a
locally underdamped response of the systems. The model given in
Eq.~(\ref{VE_model}) may be used to describe some of the effect of
topological defects near depinning, but becomes inapplicable at
large driving forces where dislocations recombine as the lattice
reorders.

Before studying the dynamics of the VE model, it is useful to
discuss its relationship to other models studied in the
literature. In particular, the form of the stress transfer
function given in Eq.~(\ref{sigmaVE}) is the same as that used
recently by Fisher and Schwarz to incorporate the effect of stress
overshoot  on propagation of cracks in heterogeneous solids
\cite{SF01,SF02}, although the random pinning force considered
there is not periodic. These authors  consider an automaton model
where time is discrete. It is straightforward, to define an
automaton version of our VE model, where both the displacement
$u_i$ and time are discrete. The displacement  takes integer
values and the automaton is updated according to the rule
\begin{eqnarray}
\label{automaton} & & u_i(t+1)=u_i+\Theta(F_i(u_i))\;,\nonumber\\
& & v_i(t+1)=\Theta(F_i(u_i))\;,
\end{eqnarray}
where  $v_i$ can have values $0$ and $1$ and $F_i(u_i)$ is the
total force at site $i$ given by
\begin{equation}
\label{force_autom} F_i(u_i)=\frac{K}{Z}\sum_{\la
j\ra}\big[u_j(t)-u_i(t)\big]+\frac{\eta}{Z(1+\eta)}\sum_{\la
j\ra}v_j(t)+F+\tilde{f}_p(i)\;.
\end{equation}
The pinning force $\tilde{f}_p(i)$ becomes a random number chosen
uniformly from an interval $[0,h_0]$ \cite{initial_cond}. The
automaton can be obtained in the limit of very deep periodic
pinning wells, when the dynamics is dominated by discrete events
corresponding to jumps of the displacement from one well to the
next. As long as $v<<1$, where 1 is the artificial upper limit of
the mean velocity in the discrete model, the automaton dynamics
mimics the continuous time dynamics reasonably well. The automaton
version of the viscoeleastic model given in Eqs.~(\ref{automaton})
and (\ref{force_autom}) is identical its dynamics to the model of
crack propagation with stress overshoot studied by Schwarz and
Fisher, provided the strength $M$ of the stress overshoot is
identified with the combination $\eta/(1+\eta)$. The two models
differ in the type of pinning considered as the random force
$f_p(i,u_i)$ used in Refs.~\cite{SF01,SF02} is not periodic. By
establishing the connection between these two models we have shown
that distinct physical mechanisms (inertia, nonlocal stress
propagation, unbound topological defects) at play in different
physical systems can be described generically by a coarse-grained
model that includes a coupling to local velocities of the driven
manifold.

In the second class of models topological defects are explicitly
allowed  by removing the constraint of single-valued
displacements.  At a strong pinning center, deformations of the
driven medium can be large and lead to the accumulation of a large
strain. When the distortion is released through a collapse of the
amplitude of the order parameter (i.e., the creation of a
topological defect), the displacement abruptly advances of an
amount of order $1$, while the amplitude quickly regenerates. This
process is known as phase slippage in superconductors and
superfluids. On time scales large compared to those of the
microscopic dynamics, it can be described approximately as a
``phase slip'': an instantaneous  (modulo $1$) hop of the
displacement of an integer unit, modeled as a coupling periodic in
the difference in displacements between neighboring degrees of
freedom,
\begin{equation}
\label{sigma_PS} \sigma^{\rm PS}(\{u_i\},t)=\frac{K}{Z}\sum_{\la
j\ra}\sin 2\pi\big[u_j(t)-u_i(t)\big]\;.
\end{equation}
If the relative displacements are small, this reduces to an
elastic coupling of strength $K$. This model has been studied
before in the mean-field limit by Strogatz and collaborators for a
sinusoidal pinning force \cite{strogatz88&89}. In this case
depinning is always hysteretic in mean field. Very recently we
were able to solve the model in mean field for arbitrary pinning
potential and show that more generically both continuous and
discontinuous depinning transitions are obtained as the parameter
$K$ is varied \cite{SSMM04}.

Below we compare the behavior of these two classes of models:
models where the displacement remains single valued and deviations
from elasticity are introduced via local inertial couplings and
models where the displacement ceases to be single valued and
topological defects can be generated for strong disorder.

\section{Mean-field solution}
The mean-field approximation for the VE model is obtained in the
limit of infinite-range elastic and viscous interactions. Each
displacement then couples to all others only through the mean
velocity, $v=N^{-1}\sum_i\dot{u}_i$, and the mean displacement,
$\overline{u}=N^{-1}\sum_iu_i$. We look for solutions moving with
stationary velocity, so that $\overline{u}=vt$. Since all
displacements $u_i$ are coupled, they can now be indexed by their
disorder parameters $\gamma$ and $h$, rather than the spatial
index $i$. The mean-field dynamics is governed by the equation
\begin{equation}
\label{MFT_viscous} (1+\eta)\dot{u}(h,\gamma)= K\big(vt-u\big)
    +F
   +\eta v+f_p(u;h,\gamma).
\end{equation}
It is useful to first review the case $\eta=0$, where
Eq.~(\ref{MFT_viscous}) reduces to the mean field theory of a
driven elastic medium worked out by Fisher and collaborators
\cite{NF92}. The mean field velocity is determined by the
self-consistency condition $\la u(h,\gamma)-vt\ra_{h,\gamma}=0$,
where the subscripts $h,\gamma$ indicate averages over the
distribution of pinning strengths, $\rho(h)$, and over the
uniformly-distributed phases, $\gamma$. For piecewise harmonic
pinning, $Y(u)=1/2-u$, for $0\leq u\leq1$, no moving solution
exists for $F<F_T=\la\frac{h^2}{2(K+h)}\ra_h$. Just above
threshold the mean velocity has a universal dependence on the
driving force, with $v\sim(F-F_T)^\beta$.  In mean-field the
critical exponent $\beta$ depends on the shape of the pinning
force: $\beta=1$ for the discontinuous piecewise harmonic force
and $\beta=3/2$ for generic smooth forces.  Using a functional RG
expansion in $4-\epsilon$ dimensions, Narayan and Fisher
\cite{NF92} showed that the discontinuous force captures a crucial
intrinsic discontinuity of the large scale, low-frequency
dynamics, giving the general result $\beta=1-\epsilon/6+{\cal
O}(\epsilon^2)$, in reasonable agreement with numerical
simulations in two and three dimensions \cite{myersSIM,aamSIM}.
For simplicity and to reflect the ``jerkiness'' of the motion in
finite-dimensional systems at low velocities, we use piecewise
harmonic pinning.
\begin{figure}[htbp]
\epsfxsize=12cm \centerline{\epsfbox{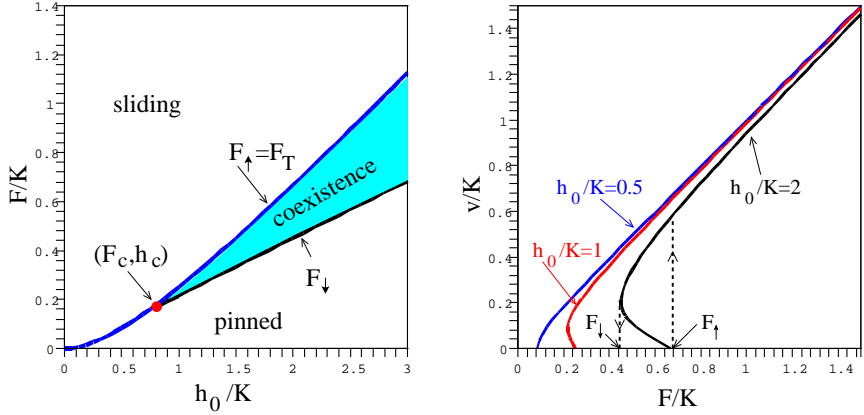}}
\caption{Mean-field solution of the VE model with a piecewise
parabolic pinning potential, $\rho(h)=\delta(h-h_0)$ and $\eta=5$.
Left frame: phase diagram. Here "coexistence" refers to
multistability of the solutions to the equations of motion. Right
frame: velocity versus drive for $h_0/K=0.5$ (blue), $h_0/K=1$
(red) and $h_0/K=2$ (black). Also shown for  $h_0/K=2$ are the
discontinuous hysteretic jumps of the velocity obtained when $F$
is ramped up and down adiabatically.} \label{fig:VE_PD}
\end{figure}


When $\eta > 0$, the nature of the depinning differs qualitatively
from the $\eta=0$ case in that both unique and multi-valued
solutions can exist  depending on the values of the parameters.
The solution for general $\eta$ can be found from that for
$\eta=0$ by substituting the effective driving force $G=F+\eta v$
for $F$ in the $v_{\eta=0}(F)$ relation and scaling the velocity
down by $1+\eta$. The linear transformation $F = G-\eta v$ then
gives the general $v(F)$ curve. The mean velocity in the sliding
state is given by the solution of
\begin{equation}
F-F_T=v\big[1-M(\eta,K)\big]+\la\frac{h^2}{K(K+h)[e^{(K+h)/(1+\eta)v}-1]}\ra_h\;,
\end{equation}
where $M(\eta,K)=(1+\eta)\la\frac{h^2}{(K+h)^2}\ra_h$. For
$M(\eta,K)<1$ there is a unique sliding solution with mean
velocity near threshold given by
\begin{equation}
\label{v_crit} v\sim\frac{(F-F_T)^\beta}{1-M(\eta,K)}\;,
\end{equation}
with $\beta=1$. The condition $M(\eta,K)=1$ determines a critical
line separating unique from multi-valued solutions $v(F)$. The
phase diagram in the $(F,h_0)$ plane is shown in
Fig.~\ref{fig:VE_PD} for $\rho(h)=\delta(h-h_0)$ (provided
$K\not=0$, the topology of the phase diagram does not depend on
the form of $\rho(h)$). There is a {\it tricritical point} at
$(h_c,F_c=F_T)$, with $h_c=K/(\sqrt{1+\eta}-1)$.  For $h_0<h_c$, a
continuous depinning transition at $F_T$ separates a stable pinned
state \cite{stabilitynote} from a sliding state with {\em unique}
velocity given by Eq.~(\ref{v_crit}).  In finite dimensions, we
expect this transition to remain in the same universality class as
the depinning of an elastic medium ($\eta=0$). This is
corroborated by numerical studies and analysis by Schwarz and
Fisher \cite{SF02} of the related stress overshoot model (but with
non-periodic pinning).  For $h_0>h_c$ The $v(F)$ curves are
multivalued, which leads to  hysteresis when $F$ is ramped up and
down adiabatically. The hysteresis is easily understood as a
consequence of the global nature of the viscous coupling in mean
field, where the driving force is replaced by an effective drive
$F+\eta v$. Clearly this does not affect the static state where
$v=0$, so that upon ramping up the driving force from the pinned
state the system always depins at $F_\uparrow=F_T$. When the force
is ramped down from the sliding state where $v\not=0$, the system
sees a larger effective drive $F+\eta v$ and repins at the lower
value $F_\downarrow$. It is important to appreciate the crucial
role of a finite value of $K$ in Eq.~(\ref{VE_model}). When $K=0$
each degree of freedom has its own velocity $v_i$ and for any
broad $\rho(h)$ there are always some degrees of freedom that
experience zero pinning force, yielding no stable pinned phase at
any $F>0$ \cite{MMP2000}. For finite long-time elasticity, i.e.,
when $K\not=0$, the elastic forces or particle conservation
enforce a uniform time-averaged velocity for all degrees of
freedom and one obtains a stable pinned phase for $F<F_T$.
Finally, we note that the VE model is also closely related to a
model of sliding CDWs that incorporates the coupling of the CDW to
normal carriers by adding a global velocity coupling
\cite{littlewood88,levy92} to the  Fukuyama-Lee-Rice model.

We now turn to the phase slip model, where the stress is given by
Eq.~(\ref{sigma_PS}). In mean field theory the nonequilibrium
state can be described in terms of two order parameters. The first
is the {\em coherence} of the phases, measured by the amplitude
$r$ of a complex order parameter defined by
\begin{equation}
\label{coherence} re^{i\psi}=\frac{1}{N}\sum_{i=1}^{N}e^{i2\pi
u_i}\;,
\end{equation}
with $\psi$ a mean phase. In the absence of interactions among the
phases or external drive, the $u_i$ are locked to the random
phases $\gamma_i$ and the state is incoherent, with $r=0$. In the
opposite limit of very strong interactions  we expect perfect
coherence of the static state, with all $u_i$ becoming equal to
the mean phase and $r\to 1$. The second order parameter is the
average velocity of the system, given by
\begin{equation}
\label{meanvel} v=\frac{1}{N}\sum_{i=1}^{N}\dot{u}_i(t)\;
\end{equation}
The mean velocity is the order parameter for the transition
between static and moving phases. In the sliding state $\psi=v t$.
The mean-field equation of motion is
\begin{equation}
{\dot u}(h,\gamma)=F - Kr \sin (2\pi u-\psi) + h Y(u-\gamma)\;,
\label{mft_eom1}
\end{equation}
where the effective  coupling is proportional to the coherence
$r$. The self-consistency condition for the mean field theory is
given by
\begin{equation}
r e^{i\psi}=\int_{0}^{1}d \gamma \int dh \rho(h) e^{i2\pi
u(h,\gamma)}\;. \label{self_consistency1}
\end{equation}

\begin{figure}[htbp]
\epsfxsize=5.5cm \centerline{\epsfbox{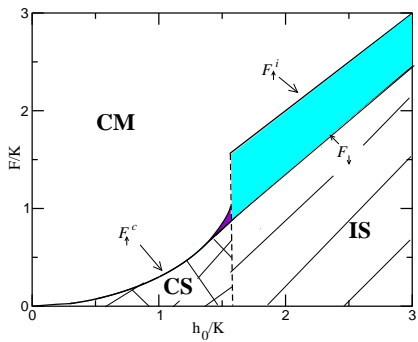}}
\caption{Mean-field phase diagram for the PS model with a
piecewise parabolic pinning potential and $\rho(h)=\delta(h-h_0)$.
In the cyan region both stable CM and IS solutions exist. In the
small purple region near the tricritical point hysteretic
depinning from the CS state is obtained numerically. The
discontinuous increase of the depinning threshold at the critical
point is a peculiarity of the piecewise pinning potential and is
replaced by a sharp, but continuous rise for other pinning
potentials.} \label{fig:PS_PD}
\end{figure}

In \cite{SSMM04} we developed a general method for solving the
mean field equations (\ref{mft_eom1}) and
(\ref{self_consistency1}) for arbitrary pinning potential. The
resulting phase diagram for a piecewise parabolic pinning
potential is shown in Fig.~\ref{fig:PS_PD}. We see immediately
that this is richer than the phase diagram for the VE model, as
now both disorder-driven transitions between static coherent and
incoherent phases as well as force-driven transitions between
static and moving phases can occur. Although, as shown in
\cite{SSMM04}, the detailed shape of the phase boundaries depends
on the specific form of the pinning potential in mean field, the
types of phases and the schematic topology of the phase diagram
are general. At low driving forces both coherent (CS) and
incoherent (IS) static phases are stable.
\begin{figure}[htbp]
\epsfxsize=5.5cm \centerline{\epsfbox{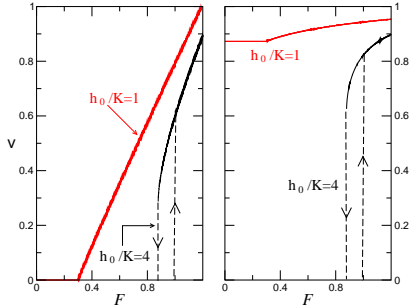}}
\caption{Coherence (right) and velocity (left) as functions of
drive for $h_0/K=1$ (red curves, showing the ${\rm CS}\rightarrow
{\rm CM}$ continuous depinning) and $h_0/K=4$ (black curves,
showing the ${\rm IS}\rightarrow {\rm CM}$ hysteretic depinning).
The curves are obtained by numerical integration of
Eq.~(\ref{mft_eom1}) as $F$ is ramped up and then down.}
\label{fig:PS_vel}
\end{figure}
The transition from the coherent phase at weak disorder to the
incoherent phase at strong disorder is discontinuous for piecewise
parabolic pinning. It also becomes hysteretic for most other
pinning potentials.\cite{SSMM04} When the driving force is ramped
up adiabatically from the coherent state, the system depins
continuously at $F_\uparrow^c$ to a unique sliding state and the
coherence grows smoothly towards unity. When the force is ramped
up from the incoherent static state, where $r=0$, the system
depins  discontinuously at $F_\uparrow^i=F_{\rm sp}$, with $F_{\rm
sp}=h_0$ the single particle threshold, given by the maximum
pinning force. This is easily understood from Eq.~(\ref{mft_eom1})
as when $r=0$ the displacements can remain decoupled even for
finite values of $K$. At $F_\uparrow^i$ the coherence also jumps
discontinuously to a finite value, as shown in
Fig.~\ref{fig:PS_vel}. When the force is ramped back down from the
sliding state, the system repins at the lower force
$F_\downarrow$, where the coherence also jumps back to zero. The
moving state is always coherent in mean field, although incoherent
sliding states should be possible in finite dimensions
\cite{olson_hyst01}. The origin of the mean-field hysteresis
observed when the system depins from a static incoherent state is
easily understood. When the force is ramped up each degree of
freedom depins essentially independently. Once the medium starts
sliding, disorder becomes less important and the system becomes
coherent and therefore much stiffer. Such a system cannot easily
adjust to disorder and therefore when the force is decreased
remains in a sliding state down to a lower force. This type of
hysteresis is clearly strongly enhanced in mean-field, where only
states with zero or perfect coherence are possible. Hysteresis has
been observed in numerical simulations of the PS model in three
dimensions \cite{Nogawa03}, although this work did not establish
conclusively that the hysteresis survives in the limit of infinite
system size. Preliminary work by us \cite{AAMunp} suggests that
there is no hysteresis in infinite systems in three dimensions.

\section{Conclusion}

We have discussed the depinning transition of two classes of
models that allow for history-dependent response. In mean field
theory both models exhibit a tricritical point as a function of
disorder strength. At weak disorder depinning is continuous and
the sliding state is unique. Above the tricritical point depinning
is discontinuous and hysteretic. Numerical studies of these models
are currently underway to establish how much of the mean field
behavior survives in finite dimensions. Simulations of the PS
model in three dimensions suggest that the hysteresis disappears
for large system sizes
\cite{Nogawa03,AAMunp,Huse1996,Kawaguchi99}. In this model the
hysteresis may indeed be an artifact of the mean field
approximation that yields a sharp distinction between coherent and
incoherent static states. In finite dimension this may be replaced
by a smooth growth of local coherence over a broad range of length
scales, without discontinuous jumps. The mean-field tricritical
point may then become  a strong crossover between continuous
depinning transitions characterized by (nonuniversal) exponents
$\beta<1$ at weak disorder to $\beta>1$ at strong disorder
\cite{Kawaguchi99}. Hysteresis seems more robust in the VE model,
or in general in models with local inertia. Preliminary numerical
simulations of an anisotropic version of the VE models in two
dimensions indicate a stubborn persistence of hysteresis with
increasing system size even for relatively weak disorder. In this
case the existence of hysteresis may not be a good test of the
precise nature of the depinning transition \cite{Maimon04}.

Once local inertia or topological defects are introduced in the
model, various depinning scenarios are possible. The depinning may
be discontinuous with hysteresis (like an equilibrium first order
phase transition) or with hysteresis that vanishes in the infinite
system limit. Another possibility is that the transition in
continuous and critical, in the sense that it is possible to
identify diverging correlation lengths as the depinning threshold
is approached adiabatically from above or from below. Finally, as
originally suggested by Ramanathan and Fisher \cite{SR_DSF98} and
more recently explored by Maimon and Schwarz \cite{Maimon04}, even
critical behavior with hysteresis that survives in the infinite
system limit is possible.   Sorting out this various scenarios for
the models discussed will require more extensive numerical studies
in finite dimensions.

\vspace{0.2in} This work is supported by the National Science
Foundation under grant DMR-0305407. Various aspects of the work
described here were carried out in collaboration with Alan
Middleton, Karl Saunders, Jen Schwarz, and Bety Rodriguez-Milla.


\begin{thebibliography}{99}
\bibitem{convective}
We drop the convective derivative as it does not contribute in
mean field.
\bibitem{stabilitynote} This stability can be history dependent
or vanish at finite driving rates, as discussed in Ref.\
\onlinecite{SF01}.
\end{thebibliography}
\end{document}